\documentclass[showpacs]{revtex4}
\usepackage{graphicx}

\begin{document}

\title{Dilute Liquid of Instanton and Its Topological Charge Dominate the QCD Vacuum}

\author{Z.Q. Wang$^{1}$}
\author{X.F. L\"{u}$^{1,3,4}$}
\author{F. Wang$^{2}$}

\affiliation{${}^{1}$Department of Physics, Sichuan University,
Chengdu  610064, China}

\affiliation{${}^{2}$Department of Physics, Nanjing University,
Nanjing 210093, China}

\affiliation{${}^{3}$Institute of Theoretical Physics, The Chinese
Academy of Sciences, Beijing  100080, China}

\affiliation{${}^{4}$CCAST (World Laboratory), P.O. Box 8730,
Beijing  100080, China}

\begin{abstract}
APE smearing and overlap-Dirac operator are combined to filter QCD
vacuum configurations. The results obtained from overlap fermions
and improved 5Li cooling are compared, both of them exhibit
structures of dilute liquid of instanton. Finally the overlap
fermions, improved 5Li cooling and APE smearing are combined to
calculate the topological charge and identify the structure of QCD
vacuum. The results suggest dilute liquid of instanton dominance of
topological charge fluctuations in quenched lattice QCD.
\end{abstract}


\

\maketitle

The notion of the gauge field topology is important for
understanding QCD vacuum. The instanton liquid model
(ILM)\cite{shur.82}, where correlation among instantons is
introduced, has been used as a basis for developing a rather
successful semiclassically-motivated phenomenology. Even though
termed as a liquid, it is quite dilute with (anti)instantons of
radius $\rho\approx 1/3$fm and density $n\approx 1$fm$^{-4}$. This
setup allows for an interesting mechanism of spontaneous chiral
symmetry breaking. Since ILM plays an important role in QCD it has
been checked by lattice QCD. It is quite reasonable to expect that
the lattice QCD will eventually provide answers about the structure
of QCD vacuum. However, finding a clean and satisfactory way to
infer this information from lattice QCD has proven to be nontrivial
since the Monte-Carlo generated gauge fields fluctuate wildly. In
order to eliminate the short-distance fluctuations it is necessary
to manipulate the gauge fields by various cooling\cite{ch.94} or
smoothing\cite{fal.85} procedures. The cooling is a local
minimization procedure for the gauge action with initial one being
the Monte-Carlo generated QCD configuration. However, after a few
cooling sweeps the gauge field undergoes large changes and become
smooth. Further cooling possibly leads to even smoother
configuration, and eventually into the trivial configuration with
nonperturbative effects removed. Therefore the cooling or smoothing
has the subjective nature since one stops at sweeps while the gauge
configuration shows the ILM structure.

Horv\'{a}th et al.\cite{hor.66} used lattice fermions to study the
vacuum structure, because the space-time structure of the low-lying
eigenmodes of overlap-Dirac operator with the Monte Carlo simulation
gauge field is naturally smoother than that of the gauge fields
themselves. They identify the possible vacuum structures by finding
the local maxima of density $d(n)=\psi_n^\dag\psi_n$. In their
criterion, the local maxima are retained only if the average of
$d(n)$ decays monotonically from origin over the distance
$\sqrt{3}a$ for all directions\cite{hor.66}. Their results exhibited
that the densities of these lump structures increase with the
lattice spacing decreasing and might be divergent in the continuous
limit. So they concluded that vacuum fluctuations of topological
charge are not effectively dominated by instantons. We suspect that
the lattice random fluctuations were not excluded in their results.
With the lattice spacing decreasing, the random lattice tiny
fluctuations become more and more. If they still used $\sqrt{3}a$ in
their criterion, which decreases with the decreasing of the lattice
spacing $a$, almost all fluctuations, real and random fluctuations,
would be included in their results. Even a real instanton superposed
by random fluctuations would be split into small lumps. In order to
get rid of this trouble, we combine APE smearing\cite{fal.85,
hasen.98, td.98}, improved 5Li cooling\cite{for.96,for.98} and
overlap-Dirac operator\cite{neu.98} methods to fix the space-time
structures of the gauge configurations and calculate the topological
charge of every structure fixed above to discriminate real and
random fluctuations. After the random fluctuations are dropped a
dilute instanton liquid QCD vacuum appears.

The APE smearing reduces the short range fluctuations effectively
but will not destroy the instantons that are larger than about 1.5
lattice spacing\cite{hasen.98}. Furthermore it accelerates the
calculation of overlap-Dirac operator approach if one performs
pretreatment of the original lattice gauge fields by a few steps of
APE smearing. We take $N=10$ levels of APE smearing with
$\alpha=0.45$ APE parameter for every configuration generated from
Iwasaki action\cite{iwa.85}. In Fig.1 we give a comparison of the
chirality structure (see below) of the lowest nonzero eigenmode of
the overlap-Dirac operator with and without pre-treatment by APE
smearing. It can be seen from Fig.1 that the structure exhibited by
the overlap-Dirac operator with and without pre-treatment are almost
the same, but the pre-treated one (Fig.1b) is smoother than that
without pre-treatment (Fig.1a). We also compared other low-lying
eigenmodes and obtained similar results. These results show that the
pre-treatment wouldn't destroy the physical structure and just
sweeps away the short range fluctuations.
\begin{figure}[htbp]
\centering
\includegraphics[width=0.4\textwidth]{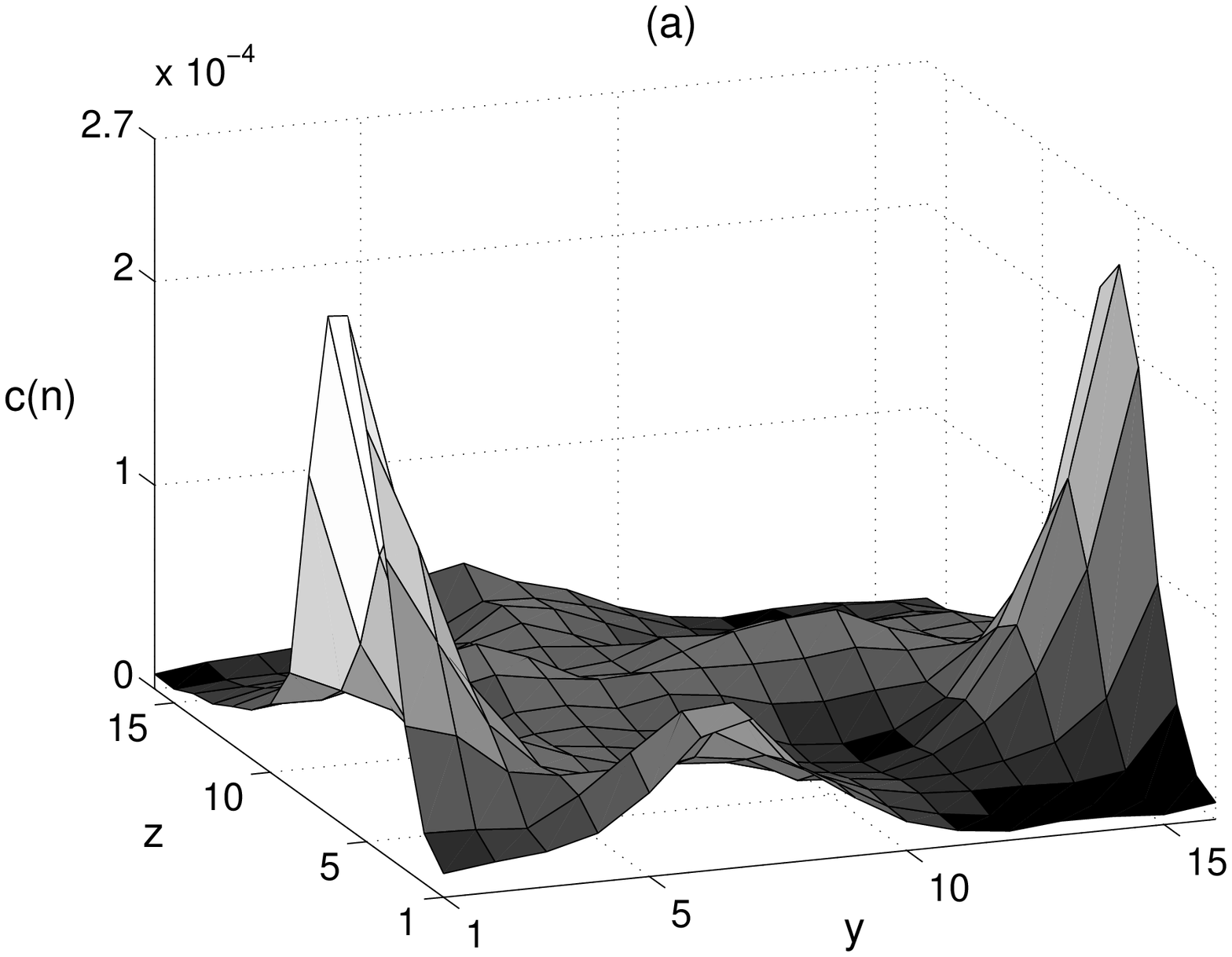}
\includegraphics[width=0.4\textwidth]{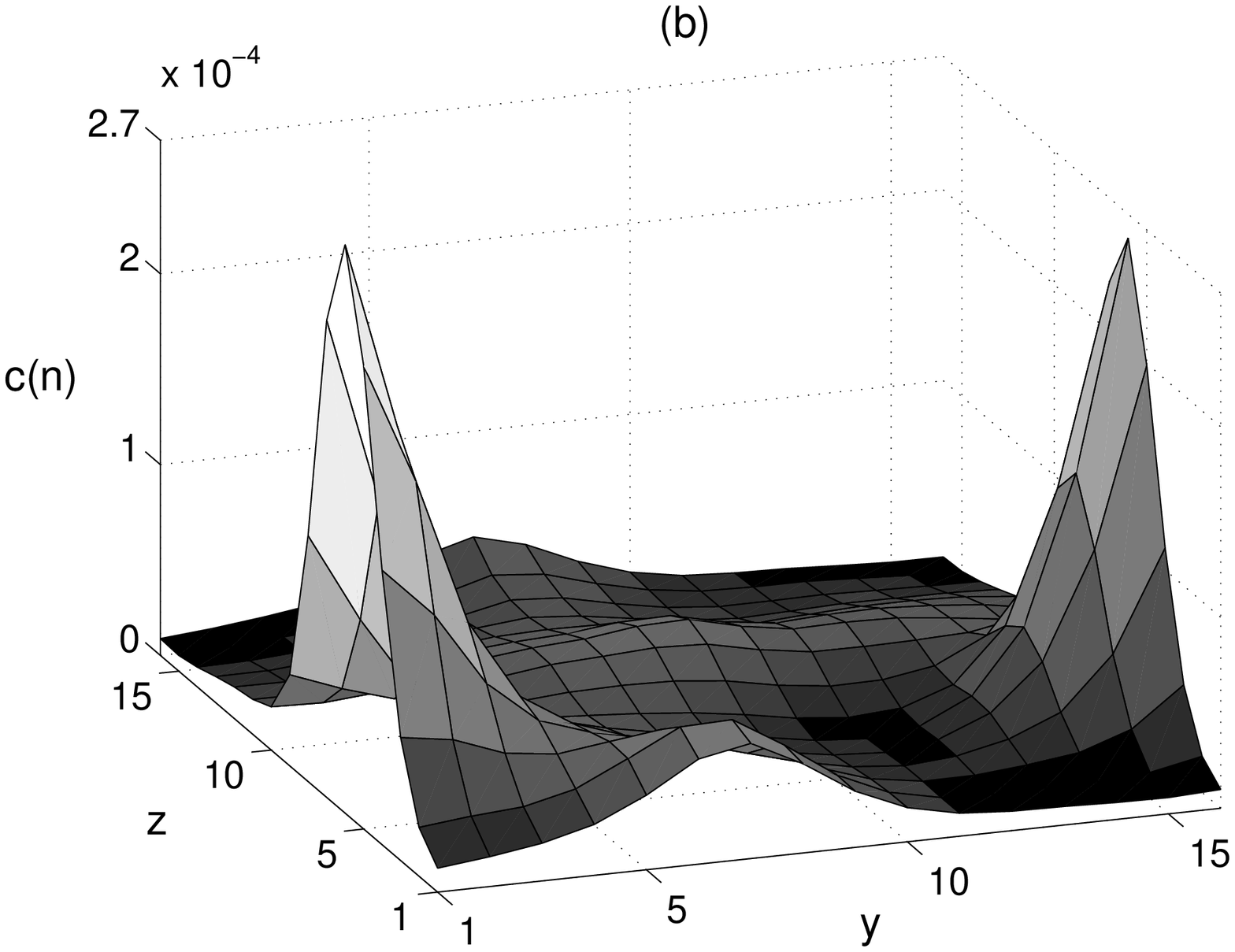}
\caption{The distribution of chirality $c(n)$ of the lowest nonzero
mode in a $y$-$z$ slice for a $16^3\times 20$ Iwasaki lattice
($\beta=2.746$, $a=0.08$fm) without(a) and with(b) APE
pre-treatment.}
\end{figure}

Our overlap-Dirac operator results show that the density
$d(n)=\psi_n^\dag\psi_n$ and the chirality
$c(n)=\psi_n^\dag\gamma_5\psi_n$ have the same space-time
structures. This is consistent with ILM . However chirality $c(n)$
is better than density $d(n)$, because using of chirality one can
distinguish the structures with different chiral property of QCD
vacuum. From our overlap-Dirac operator results with APE pre-treated
gauge configurations we select out those points which have the
biggest chirality and the smallest chirality separately and examine
whether they condensate as lumps. Both the zero modes and the lowest
nonzero modes of the overlap-Dirac operator are analyzed. As showed
in Table 1, different sample of points have been selected to explore
the details of structures for a $16^3\times20$ lattice. We found
that there are a few lumps in the zero mode and the lowest nonzero
mode and that when the number of points selected increases some
small lumps merge into bigger ones. In the following we will show
that the small lumps remained can not be regarded as
(anti-)instantons, since topological charge of these structures is
far smaller than one; while the bigger lumps have topological charge
near the topology integer 1. We also show the distribution of
chirality of the lowest nonzero mode in a given time slice in
Fig.2(a). This figure exhibits perfect structures of the low-lying
mode, which hints that the background gauge field has space-time
structures.
\begin{table}[htbp]
 \caption{{\scriptsize The chirality structures of the zero mode and
the lowest nonzero mode of a $16^3\times20$ lattice at
$\beta=2.746$. The first sub-table lists the structures of the zero
mode which is positive chirality. The second one lists the
structures of the lowest nonzero mode with positive chirality. The
last one lists the structures of the lowest nonzero mode with
negative chirality. In every sub-table, the first line lists the
numbers of points selected and the value of chirality $c(n)$ on the
iso-surfaces (listed in brackets); the lower lines list the
coordinates of the centers of every lump and the numbers of points
used (listed in brackets) for every lump.}}
\vspace*{0.5cm}
{\scriptsize
\begin{tabular}{l|l|l|l|l}\hline\hline
\multicolumn{5}{l}{The structures of the Zero mode (positive
chirality):}\\  \hline
  64 ($3.50\times10^{-4}$ )& 128 ($2.84\times10^{-4}$ )& 512 ($1.48\times10^{-4}$ )&
  1024 ($9.98\times10^{-5}$ )& 2048 ($6.61\times10^{-5}$ )\\  \hline
  12.7 15.1 12.2 12.3 (64)&12.6 15.2 12.3 12.2 (128)&12.5 15.4 12.5 12.2 (512)&
  12.2 15.3 12.3 12.0 (1018)&12.5 15.4 12.5 12.2 (2018)\\
 & & &4.6~~~0.5~~~7.0~~7.3~~(6)&4.8~~~0.5~~~6.7~~7.5~~(30)\\  \hline
\multicolumn{5}{l}{The structures of the lowest nonzero mode with
positive chirality:} \\ \hline
  64 ($1.39\times10^{-4}$ )& 128 ($1.06\times10^{-4}$ )& 512 ($6.74\times10^{-5}$ )&
  1024 ($5.33\times10^{-5}$ )& 2048 ($4.16\times10^{-5}$ )\\  \hline
  5.0~~~0.9~~~5.9~~7.7~~(64)&5.1~~~0.8~~~6.0~~7.7~~(128)&5.2~~~1.1~~~5.8~~7.9~~(304)&
  5.1~~~1.6~~~5.5~~8.1~~(496)& \\
   & &4.0~~~6.5~~~2.0~~8.5~~(4)& &4.3~~~3.3~~6.0~~10.1 (1517)\\
   &3.4~~~3.7~~7.1~~14.7~~(7)&3.4~~~4.0~~6.9~~14.3 (101)&3.2~~~4.3~~7.2~~13.3
   (283)& \\
   & &2.8~~~5.4~~8.3~~9.6~~~(9)& & \\
   &10.0 10.0 10.0 9.5~~(2)&10.2 10.3 10.1 9.5~~(57)&10.1 10.4 10.2
   9.5~~(139)&\\
   &&12.8 14.6 12.2 12.2 (37)&12.7 14.7 12.1 12.2 (106)&11.2 12.4 11.0 10.6
   (531)\\ \hline
\multicolumn{5}{l}{The structures of the lowest nonzero mode with
negative chirality:} \\ \hline
  64 ($-2.10\times10^{-4}$ )& 128 ($-1.69\times10^{-4}$ )& 512 ($-9.23\times10^{-5}$ )&
  1024 ($-6.77\times10^{-5}$ )& 2048 ($-4.87\times10^{-5}$ )\\  \hline
  1.1~~~6.9~~15.7 0.6~~(64)&1.2~~~7.0~~15.7 0.6~~(128)&1.3~~~6.8~~15.3 0.5~~(512)&
  1.4~~~6.8~~14.8 0.6~~(1024)&1.7~~~6.6~~15.2 0.6~~(2048)\\ \hline\hline

\end{tabular}
}
\end {table}

Then a comparison of the results obtained from overlap fermions and
improved 5Li cooling is made. The improved 5Li cooling\cite{for.96,
for.98} is a local minimization of the following 5Li lattice action
\begin{equation}
S=\sum_{i=1}^5c_iS_{m_i,n_i},
\end{equation}

\begin{equation}
S_{m,n}=\frac{1}{m^2n^2}\sum_{x,\mu,\nu}Tr\left[1-\frac{1}{2}\left(R_{x,\mu
\nu}^{m\times n}+R_{x,\mu \nu}^{n\times m}\right)\right],
\end{equation}
where $R_{x,\mu \nu}^{m\times n}$ denotes the ordered product of
SU(3) link matrices along $m\times n$ rectangles in the $\mu$, $\nu$
plane, $(m_i,n_i)$=(1,1),(2,2),(1,2),(1,3),(3,3) for $i=1,...,5$ and
$c_1=(19-55c_5)/9$, $c_2=(1-64c_5)/9$, $c_3=(-64+640c_5)/45$,
$c_4=1/5-2c_5$, $c_5=1/20$. The general cooling with Wilson action
has no stable instantons, even it existed in the gauge field, due to
shrink and decay under cooling. But the 5Li cooling overcomes this
difficulty to some extent. The 5Li action provides a small energy
barrier against the decay of instantons of size $\rho\ge2.3a$, thus
preserves instantons over that size under arbitrary amounts of
cooling\cite{for.98}. After cooling, one can use topological charge
density operator\cite{ch.94}
\begin{equation}
q(x)=-\frac{1}{32\pi^2}\epsilon_{\mu \nu \rho \sigma}ReTr[F_{\mu
\nu}(x)F_{\rho \sigma}(x)]
\end{equation}
to measure the topological charge distribution of the cooled
configurations. Although this operator is inadequate for
configurations of large fluctuations, it is adequate for the smooth
configurations which emerge after several cooling steps. We found
that the improved 5Li cooling and the overlap fermions (with APE
smeared background gauge field) can generate similar structures if
we adjust cooling steps. In Fig.2 we compared the 5Li cooling and
overlap fermions results. Fig.2(a) is the chirality distribution of
the lowest nonzero eigenmode of overlap-Dirac operator in a given
time slice ($t=9$) for a $16^3\times20$ lattice generated with the
Iwasaki gauge action\cite{iwa.85} at $\beta=2.746$(spacing
$a=0.08$fm). Fig.2(b) is the topological charge density distribution
of cooled configuration with 15 5Li cooling steps for the same
lattice. From Fig.2 one can see that 5Li cooling and overlap
fermions filter out almost the same structure but the two methods
differ from each other significantly, which suggests that the
topological information thus obtained is physical. So these two
methods can be combined to study the vacuum structure.
\begin{figure}[htbp]
\centering
\includegraphics[width=0.4\textwidth]{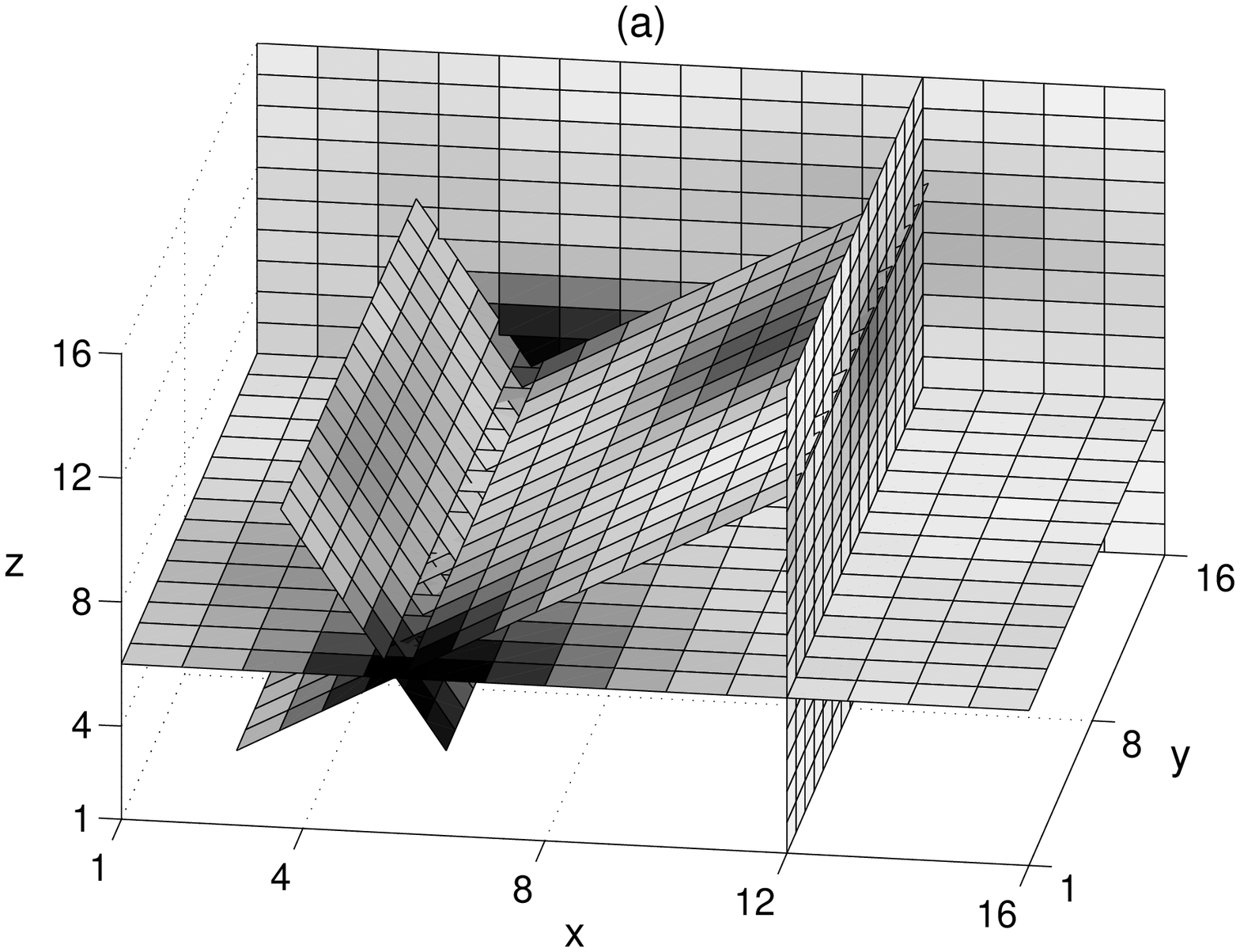}
\includegraphics[width=0.4\textwidth]{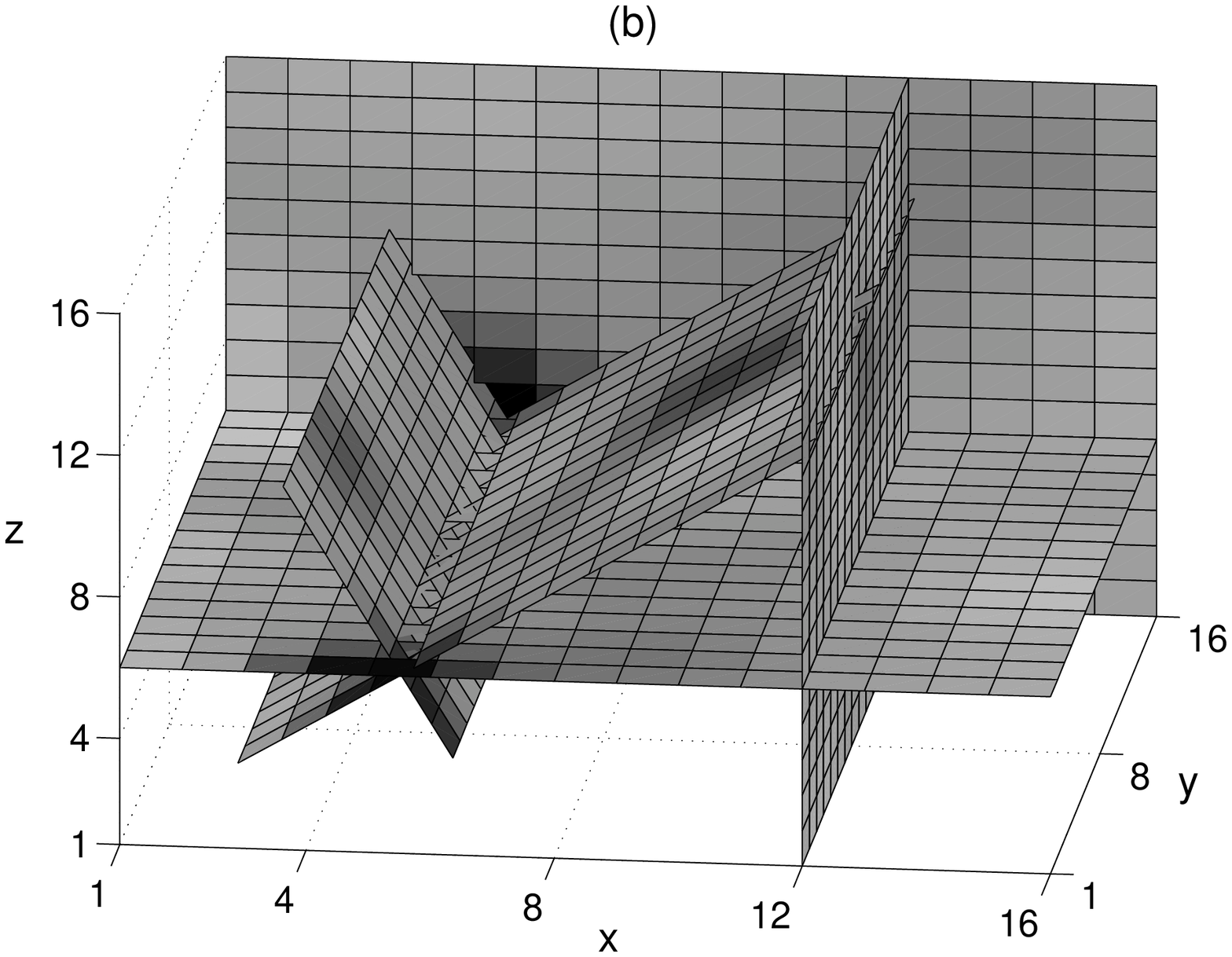}
\caption{The chirality $c(n)$ distribution of the lowest nonzero
eigenmodes of overlap-Dirac operator (a) and the topological charge
density distribution of 5Li cooled configuration(b) in a given time
($t=9$) slice of a $16^3\times 20$ lattice at $\beta=2.746$ (spacing
$a=0.08$fm). Periodic boundary conditions are imposed. The darkness
denotes high chirality and high topological charge density
separately.}
\end{figure}

Moreover we found that if we take about 50 APE smearing steps all
these three filter methods can produce similar results. In Fig.3 we
showed the topological charge distribution of smoothed configuration
(with 50 APE smearing steps) for the same lattice of Fig.2.
\begin{figure}[htbp]
\centering
\includegraphics[width=0.4\textwidth]{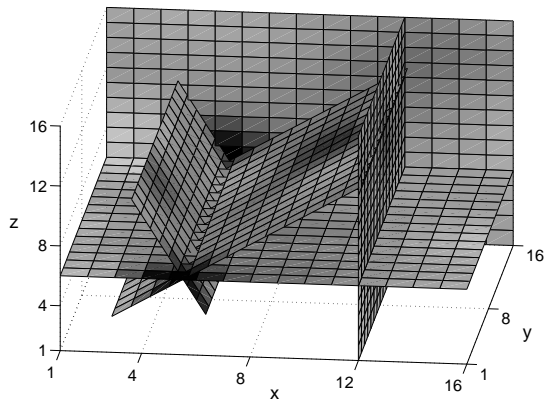}
\caption{The topological charge distribution of smoothed
configuration (with 50 APE smearing steps) for the same lattice of
Fig.2. The darkness denotes high topological charge density.}
\end{figure}

Now we can study further the topological charge of the structures
of the vacuum. The topological charge $Q$ of a structure within
the range of radius $r$ can be calculated by
\begin{equation}
Q(r)=\sum_{|x-x_0| \leq r}q(x),
\end{equation}
where $x_0$ is the center of the structure. If $|Q(r)|$ reaches
0.4, this structure may be regarded as an (anti-)instanton, and
$r$ can be regarded as its radius approximately, since the
analytical instanton topological charge density is
\begin{equation}
Q_{\rho}(x)=\frac{6}{\pi^2\rho^4}
\left[\frac{\rho^2}{(x-x_0)^2+\rho^2}\right]^4,
\end{equation}
where $\rho$ is the radius parameter, and the topological charge
contribution within radius $\rho$ of an instanton is 0.5. Sometimes
two structures are close each other. If the two structures have the
total topological charge which is smaller than one and greater than
0.4, they are considered as one structure; if their total
topological charge greater than one the two structures are
considered to be independent ones.

With above criterion, we get our results showed in Table 2. Our
calculation is done on a $16^3\times20$ lattice with 8 gauge
configurations generated with the Iwasaki gauge action\cite{iwa.85}
at $\beta=2.746$ (spacing $a=0.08$fm, the physical volume is
3.36fm$^4$) and cooled by improved 5Li method. From Table 2 we can
see that for every configuration the total topological charge near
an integer and is consistent with the number of zero modes, which
are consistent with ILM. The mean density of structures is
1.23fm$^{-4}$ and their mean radius is $\rho\approx0.395$fm, which
is a little bigger than those of ILM.

If we associate the overlap fermions with APE smearing, we can
obtain the same result as above, i.e., there are the same number of
structures for every configuration. The mean radius of instantons is
$\rho\approx0.397$fm. The total topological charge for every APE
smeared configurations (with 50 APE steps) are listed in Table 2
also, which are close to the results of the improved 5Li cooled
configurations. This confirms the result furthermore.
\begin{table}[htbp]
\caption{The number of zero modes, total topological charge and the
number of structures for every configuration. For the total
topological charge, the upper line calculated from 5Li cooled
configurations and the lower line calculated from APE smeared
configurations. } \vspace*{0.5cm}
\begin{center}
\begin{tabular}{c|c|c|c|c|c|c|c|c}\hline \hline
configuration & 1 & 2 &  3 & 4 & 5 & 6 & 7 & 8\\ \hline \# of zero
modes & 1 & 3 &  1 & 1 & 0 & 3 & 1& 1\\ \hline
total topological & $-0.88$ & 2.75 &  0.87 & $-0.95$ & 0.06 & 2.76 & $-0.99$& 0.97\\
  \cline{2-9}
charge  & $-0.87$ & 2.75 &  0.86 & $-0.97$ & $-0.04$ & 2.75 &
$-1.01$& 0.99\\ \hline
 \# of  structures  & 5 & 7 &  5 & 5 & 2 & 3
& 3 & 3\\ \hline \hline

\end{tabular}
\end{center}
\end{table}

Our results show that even after APE smearing or 5Li cooling there
are still short range fluctuations with very small topological
charges. This confirms our suspicion that there are random
fluctuation lump structures included in Horv\'{a}th et al.'s
results. If these random short wave length fluctuations are dropped,
a dilute instanton liquid QCD vacuum structure appears. Based on
these results we concluded that the dilute instanton liquid
dominates the QCD vacuum. However, since the lattice volume is only
3.36fm$^4$, the random fluctuations might be very serious. In
addition we have to decrease the lattice spacing $a$ to check if all
of these results are stable against the lattice spacing changes.
Whether the large quantum fluctuation of the QCD vacuum conjectured
by Witten\cite{wit.79} will destroy the instanton structure, our
results show not, but needs to be studied further as well.

This project supported in part by the NSFC under Grant No. 90103018,
90503011.

\end{document}